\documentclass[useAMS,usenatbib,a4paper,preprint]{mn2e}
\usepackage{graphicx}
\usepackage{amsmath}
\usepackage{txfonts}
\usepackage{mathrsfs}
\usepackage{color}

\def\aap{A\&A}
\def\apjl{ApJ}

\def\apjs{ApJS}

\newcommand{\be}{\begin{equation}}
\newcommand{\ee}{\end{equation}}
\newcommand{\bea}{\begin{eqnarray}}
\newcommand{\eea}{\end{eqnarray}}

\title[High-$z$ Quasars]{Cosmological test using the Hubble diagram of high-$z$ quasars}
\author[Melia]{Fulvio Melia\thanks{John Woodruff Simpson Fellow} \\
        Department of Physics, The Applied Math Program, and Department of Astronomy,
        The University of Arizona, AZ 85721, USA; \\ E-mail: fmelia@email.arizona.edu}

\begin{document}

\date{}

\pagerange{\pageref{firstpage}--\pageref{lastpage}} \pubyear{2019}

\maketitle

\label{firstpage}

\begin{abstract} 
It has been known for over three decades that the monochromatic
X-ray and UV luminosities in quasars are correlated, though non-linearly.
This offers the possibility of using high-$z$ quasars as standard
candles for cosmological testing. In this paper, we use a recently
assembled, high-quality catalog of 1598 quasars extending all the
way to redshift $\sim 6$, to compare the predictions of the $R_{\rm h}=ct$
and $\Lambda$CDM cosmologies. In so doing, we affirm that the parameters
characterizing the correlation depend only weakly on the chosen cosmology,
and that both models account very well for the data. Unlike $\Lambda$CDM,
however, the $R_{\rm h}=ct$ model has no free parameters for this work, 
so the Bayesian Information Criterion favours it over $\Lambda$CDM with a 
relative likelihood of $\sim 88\%$ versus $\sim 10\%$. This 
result is consistent with the outcome of other comparative tests, many of 
which have shown that $R_{\rm h}=ct$ is favoured over the standard model 
based on a diverse range of observations.
\end{abstract}


\section{Introduction}
The discovery of quasars at redshifts $z>5-6$ (Fan et al. 2003; Jiang et al. 2007; 
Willott et al. 2007; Mortlock et al. 2011; Banados et al. 2014) represents---on 
the one hand---an enduring mystery in astronomy given that the emergence of 
$10^{9-10}\;M_\odot$ supermassive black holes only $\sim 900$ Myr after the
big bang and, worse, only $\sim 500$ Myr after their likely seeding in Pop II
stellar explosions (Melia 2013a), is inconsistent with the timeline in $\Lambda$CDM.
To address this problem, one needs to invoke either an anomalously high accretion 
rate (Volonteri \& Rees 2006) or the generation of unusually massive seeds (Yoo 
\& Miralda-Escud\'e 2004). But neither of these effects has ever been observed. 
For example, Willott et al. (2010) found that no known high-$z$ quasar accretes 
at more than $1-2$ times the Eddington limit (see also Melia 2014; Melia \& 
McClintock 2015).

On the other hand, these bright objects are being studied at redshifts well beyond
the current reach of Type Ia SN observations, which tend to be restricted to redshifts
$\lesssim 2$ (see, e.g., Jones et al. 2013). They therefore provide a crucial 
probe of the geometry of the Universe at luminosity distances not easily accessible by 
other means. In recent years, a Hubble diagram of high-$z$ quasars has been constructed 
using various techniques for the identification of standard candles, though with variable
success depending, in large part, on the quality and size of the sample.

For example, the Mg II FWHM and UV luminosity of quasars at $z>6$ appear to be
correlated, and since reverberation mapping of their broad lines can at once
yield the optical/UV luminosity and the distance from the central ionizing source 
to the line-emitting gas, these observations allow us to measure the black-hole 
mass (Blandford \& McKee 1982; Kaspi et al. 2000; Bentz et al. 2009; Wandel et al. 1999;
Shen et al. 2008; Steinhardt \& Elvis 2010). And since these high-$z$ quasars also
appear to be accreting near their Eddington limit, these measurements may thus be used to
construct a Hubble diagram. Unfortunately, the sample ($\sim 20$ sources) available for 
this kind of work is still modest, so the outcome of this analysis is suggestive,
though not yet conclusive (Melia 2014). Nonetheless, the analysis based on this technique
showed, in a one-on-one comparison, that the Friedmann-Lema\^itre-Robertson-Walker (FLRW)
cosmology, known as the $R_{\rm  h}=ct$ universe (Melia 2003, 2007, 2013b; Melia \& Shevchuk 
2012) is preferred by these data over the standard model $\Lambda$CDM. Statistical tools, 
such as the Akaike (1973), Kullback, and Bayes (Schwarz 1978) Information Criteria, 
all demonstrated that the former model is favoured over the latter with a likelihood of 
$\sim 85\%$ versus $\sim 15\%$. An analogous study using a sample of 35 active galactic
nuclei at lower redshifts produced very similar, confirming results (Melia 2015a).

A very different technique, based on the flat spectrum radio quasar (FSRQ) gamma-ray luminosity 
function (Zeng et al. 2016), produced much stronger conclusions, mainly because of the much bigger
sample of such sources assembled with {\it Fermi} during its four-year survey. Using a
Kolmogorov-Smirnov test on one-parameter cumulative distributions in luminosity, redshift,
photon index and source count, this study concluded that $R_{\rm h}=ct$ is very strongly
favoured over $\Lambda$CDM. 

A third approach (L\'opez-Corredoira et al. 2018), much more closely aligned with the work reported 
in this paper, uses a recently refined method of sampling the redshift-distance relationship 
(Risaliti \& Lusso 2015), based on a correlation between the X-ray and UV monochromatic 
luminosities of quasars, first discussed in this context over three decades ago by Avni 
\& Tananbaum (1986). The fact that this relationship is seemingly
independent of evolution may render this method the best used with quasars thus far. In
the comparative analysis of nine different cosmological models carried out by 
L\'opez-Corredoira et al. (2018), only two showed very strong consistency with these
quasar data, while several others, including Einstein-de Sitter, the Milne universe
and static universe models based on tired light, were excluded at $>99\%$ confidence level.

In this paper, we revist the X-ray versus UV luminosity correlation for quasars, in light
of the much improved sample of suitable sources assembled from the literature and follow-up 
observations by Risaliti and Lusso (2019). These data are interpreted in the context
of a generally accepted scenario, in which the UV photons are emitted by an accretion
disk, while the X-rays are Compton upscattered photons from an overlying, hot corona.
But though a non-linear correlation to link these two spectral features has been known
for many years, only recently has an uncomfortably large dispersion in the relation been 
mitigated with a more refined selection of the sources. Most of the
quasars in the parent sample have been identified from the cross-correlation of the
XMM-{\it Newton} Serendipitous Source Catalogue Data Release 7 (Rosen et al. 2016)
with the Sloan Digital Sky Survey (SDSS) quasar catalogues from Data Release 7
(Shen et al. 2011) and 12 (Paris et al. 2017). With the aim of whittling this
catalog down to those sources with reliable measurements of the intrinsic
X-ray and UV emissions, avoiding possible contaminants and unknown systematics, 
Risaliti and Lusso (2019) produced a final, high-quality catalog of 1598 quasars
that we shall use for the analysis reported in this paper.

In \S~II, we introduce the two cosmological models we shall compare in this study,
along with a `cosmographic' empirical fit used by Risaliti \& Lusso (2019) to
provide a less-model dependent analysis of the quasar data. Ironically, this 
approach ends up requiring more unknown parameters for the fit than the models themselves, 
so it actually does not fare better than the latter when used with Information Criteria.
We discuss the results of our model comparison in \S~III, and conclude in \S~IV.

\section{Methodology}
The non-linear relation between the quasar's UV (disk) and X-ray (coronal)
emissions is usually parametrized as
\begin{equation}
\log_{10} L_X=\gamma\log_{10} L_{UV}+\beta\;,
\end{equation}
where $L_X$ and $L_{UV}$ are the rest-frame monochromatic luminosities
at 2 keV and 2,500 \AA, respectively. The logarithms in this expression may be
taken in any base, though we restrict our usage to base 10 throughout
this paper. Previous studies have shown that $\gamma\sim 0.5-0.7$
(Avni \& Tananbaum 1986; Just et al. 2007; Young et al. 2010; Lusso et al. 
2010; Risaliti \& Lusso 2015; Lusso et al. 2016; Risaliti \ Lusso 2019).
Since the data contain the fluxes, rather than the absolute (model-dependent)
luminosities, we use Equation~(1) in the modfied form
\begin{equation}
\log_{10}F_X=\tilde{\beta}+\gamma\log_{10} F_{UV}+2(\gamma-1)\log_{10} d_L\;,
\end{equation}
where $\tilde{\beta}$ is a constant that subsumes the slope $\gamma$ and
the intercept $\beta$ in Equation~(1), such that
\begin{equation}
\tilde{\beta}=\beta+(\gamma-1)\log_{10} 4\pi\;,
\end{equation}
 
For each model, we optimize its parameters (if it has any), along with those
characterizing the empirical fit in Equation~(2), i.e., $\gamma$, $\tilde{\beta}$
and $\delta$ (see below), by minimizing the likelihood function
\begin{equation}
\ln(LF)=-\sum_{i=1}^{1598}\left\{{\left[\log_{10}\left(F_X\right)_i-\Phi\left(
\left[F_{UV}\right]_i,d_L\left[z_i\right]\right)\right]^2\over{\tilde{\sigma_i}}^2} +
\ln\left({\tilde{\sigma_i}}^2\right)\right\}\;,
\end{equation}
where the variance ${\tilde{\sigma_i}}^2\equiv\delta^2+\sigma_i^2$ is given in 
terms of the global intrinsic dispersion, $\delta$, and the measurement error 
$\sigma_i$ in $(F_X)_i$ (Risaliti \& Lusso 2015). The error in $(F_{UV})_i$ is presumed 
to be insignificant compared to $\sigma_i$ and $\delta$, and is therefore ignored in 
this application.  In Equation~(4), the function $\Phi$ is defined as
\begin{equation}
\Phi\left(\left[F_{UV}\right]_i,d_L\left[z_i\right]\right)\equiv\tilde{\beta}+
\gamma\log_{10}\left(F_{UV}\right)_i+2(\gamma-1)\log_{10} d_L(z_i)\;,
\end{equation}
in terms of the measured fluxes $(F_X)_i$ and $(F_{UV})_i$ at redshift $z_i$.

The three models we compare here are (i) flat $\Lambda$CDM, 
(ii) the $R_{\rm h}=ct$ universe, and (iii) a cosmographic (empirical) fit
using a low-order polynomial for the luminosity distance. We have included
$R_{\rm h}=ct$ because a variety of previously published comparative tests
between this model and $\Lambda$CDM have shown that it accounts for the
data as well, if not better, than the standard model. Some counter claims 
have also been published in recent years, and we summarize some of these
here to demonstrate that the issue of whether or not $R_{\rm h}=ct$ is
the correct cosmology still needs to be resolved. The analysis in this paper
advances this discussion significantly by providing an important new
comparison between these models using observations over an unprecedentedly
large range in redshifts (see fig.~5 below).

Over the past decade, $R_{\rm h}=ct$ has been compared to $\Lambda$CDM using
data at low, high, and intermediate redshifts, based on a variety of integrated 
observational signatures, such as the luminosity and angular-diameter distances,
and also differential measures, such as the redshift-dependent Hubble parameter.
A very helpful summary of these analyses and their outcomes appears in Table~2 
of Melia (2018a). Some of these data, however, are often subject to
unknown systematics and, worse, are themselves sometimes dependent on the
assumed background model. A well-known example is the analysis of Type Ia
SNe, whose lightcurve is characterized by 3 or 4 so-called `nuisance'
parameters that must be optimized along with the unknown variables in the
cosmological model. Depending on the assumptions made (e.g., with regard
to unknown intrinsic dispersions) and techniques used (e.g., $\chi^2$
minimization versus the maximization of the likelihood function and its
relation to information criteria), the outcome of these tests can sometimes
change from one application to another. 

So in addition to the results reported in Table~2 of Melia (2018a), some
counterclaims have also been published, questioning whether $R_{\rm h}=ct$ 
is in fact favoured by the data over $\Lambda$CDM. This is the principal
reason why additional comparative tests, such as that presented in this
paper, are critical to this ongoing discussion. For example, the use of 
Type Ia SNe for model testing is especially difficult when merging 
disparate subsamples to improve the statistics, since each comes with 
its own unknown systematics. Shafer (2015) combined the Union2.1 and 
JLA samples for this purpose, and concluded that these data favour the
standard model. But in his analysis, he avoided having to deal with
the unknown intrinsic dispersions by constraining the reduced $\chi^2$
of each subsample to be 1. A better statistical approach (Kim 2011;
Wei et al. 2015; Melia et al. 2018e) estimates these unknowns by instead 
maximizing the overall likelihood function. Depending on which of these
assumptions and methods one adopts, the outcome of which model is 
preferred by the SN data changes. 

A more recent example is the use of local probes for model testing
(Lin et al. 2018), in which SN data were combined with measurements
of $H(z)$ and baryon acoustic oscillations (BAO). These authors also
concluded that $\Lambda$CDM is favoured by these observations, in 
contrast to other work in which the opposite result was reported 
(see, e.g., Melia \& L\'opez-Corredoira 2017; Melia et al. 2018e).
The difference can be easily traced to which data sets were used
in the two studies. Like the SN measurements, the BAO themselves
generally do not provide model-independent information because
the BAO peak location cannot be disentangled from redshift space 
distortions. To date, only 3 or 4 BAO measurements have provided
an unambiguous peak location. Thus, any use of BAO data, and
measurements of $H(z)$ derived from them, yields a biased
outcome. Lin et al. (2018) used all of the data and, not surprisingly,
concluded that $\Lambda$CDM is favoured, because the standard model
was assumed as the background in order to estimate the redshift 
space distortions. But when only the model-independent data are 
used (see, e.g., Melia \& L\'opez-Corredoira 2017) one reaches
the opposite conclusion.

This is a principal reason why our model-independent approach in
this paper is indispensible. To fully utilize the formalism developed
in Equations~(1)-(5), we need the luminosity distances in the three 
models we consider here, which are given as
\begin{equation}
d_L^{R_{\rm  h}=ct}(z)={c\over H_0}(1+z)\ln(1+z)\;;
\end{equation}
\begin{equation}
d_L^{\Lambda{\rm CDM}}={c\over H_0}(1+z)\int_0^z{du\over
\sqrt{\Omega_{\rm m}(1+u)^3+\Omega_\Lambda}}
\end{equation}
for the minimal 1-parameter flat concordance model, in which $\Omega_{\rm m}$
is the scaled matter density today and the cosmological constant energy density is 
$\Omega_\Lambda=1-\Omega_{\rm m}$; and
\begin{equation}
d_L^{\,\rm cos}=\ln(10){c\over H_0}\left[\log_{10}(1+z)+a_2\log_{10}^2(1+z)
+a_3\log_{10}^3(1+z)\right]
\end{equation}
for the ``cosmographic'' empirical fit, based on a third-order polynomial with
two constants, $a_2$ and $a_3$, that need to be optimized along with the other
free parameters (see Risaliti \& Lusso 2019).

The Hubble constant $H_0$ is not independent of $\tilde{\beta}$, and is subsumed
into this parameter during the optimization procedure. To make the figures easy
to interpret, however, particularly the Hubble diagram, we assume a fiducial value
$H_0=70$ km s$^{-1}$ Mpc$^{-1}$ throughout this paper. If one prefers
a different Hubble parameter, say $H_0^\prime$ then, according to Equation~(2), the 
optimized values of $\tilde{\beta}$ in Table~1 simply need to be changed by an amount
$\Delta{\tilde{\beta}}=2(\gamma-1)\log_{10}(H_0^\prime/H_0)$. Notice, therefore, that 
in this model comparison, the $R_{\rm h}=ct$ cosmology has no free parameters, 
$\Lambda$CDM has one, and the cosmographic fit has two. This difference is crucial
when evaluating the Bayesian Information Criterion (BIC) for the comparison of the
model fits, as we shall see shortly. 

When models being compared have different numbers of free parameters, as is
the case for $R_{\rm h}=ct$ and $\Lambda$CDM, a simple $\chi^2$-minimization,
or even a comparison of likelihood functions, is not sufficient to fairly
decide which cosmology is a better match to the data. The sample we have
here is very large, and therefore the most appropriate model selection tool
to use is the Bayes Information Criterion (Melia \& Maier 2013), which 
approximates the computation of the (logarithm of the) `Bayes factor' for 
deciding between models (Schwarz 1978; Kass \& Raftery 1995). In general
terms, information criteria such as this may be viewed as enhanced `goodness of 
fit' tests, extending the better known $\chi^2$ criterion by incorporating
the number of model parameters. Information criteria penalize models with 
a larger number of unknowns, unless they yield a substantially better 
fit to the data. This enhancement reduces the possibility of overfitting,
arising from the fact that, with more parameters, one may simply be adjusting
to the noise.

For each model being tested, the BIC is defined by
\begin{equation}
\label{eq:7}
\exp(-{\rm BIC}/2)\equiv n^{-k/2} L^*\;,
\end{equation}
where $L^*=e^{-\ln(LF)}$ is the maximized likelihood (from Eq.~4), $n$ ($=1598$ here) 
the data set size, and $k$~the count of free parameters in the model. Then,
the {\it relative} likelihood of model $\alpha$ being correct is the Bayes probability 
\begin{equation}
P(\alpha)= \frac{\exp(-{\rm BIC}_\alpha/2)}
{\sum_\beta \exp(-{\rm BIC}_\beta/2)}\;,
\end{equation}
where $\exp(-{\rm BIC}_\alpha/2)$ is its `Bayes weight.'
The Bayesian interpretation of this method is as follows:
$\exp\left(-{\rm BIC}_\alpha/2\right)$ is a large-sample
($n\to\infty$) approximation to an integral over the parameter
space of model~$\alpha$, of its likelihood function~$L$. The 
standard error of each parameter shrinks like~$n^{-1/2}$ in this limit.

These probabilities are, of course, relative ones so, 
in principle, the ratio of any two of them represents the factor by
which either model is more likely to be ``closer to the correct"
cosmology. For the BIC, there is an accepted interpretation of the
magnitude of the difference $\Delta={\rm BIC}_2-{\rm BIC}_1$ in
terms of the strength of the evidence against model 2 (Kass
\& Raftery 1995; Tan \& Biswas 2012). The rule of thumb is that if 
$\Delta\lesssim 2$, the evidence is weak; if $\Delta\sim 3-4$, it is 
ajudged to be mildly strong; and if $\Delta\gtrsim 5$, it is quite strong.
As we shall see, in all the cases we consider in this paper, $\Delta\sim 4$
for each pair of models being compared, so the evidence would be judged
`positive,' meaning that it is on the borderline between mildly strong and 
quite strong.

\begin{table*}
\begin{center}
\caption{Model comparison based on high-$z$ quasars}
\begin{tabular}{lccccccccc}
&&&&&&&&& \\
\hline\hline
&&&&&&&&& \\
Model& $\tilde{\beta}$ & $\gamma$ & $\delta$ & $\Omega_{\rm m}$ & $a_2$ & $a_3$ & $\chi^2_{\rm dof}$ & BIC & Likelihood  \\
&&&&&&&&& \\
\hline
&&&&&&&&& \\
$R_{\rm h}=ct$ & $6.618\pm0.011$ & $0.640\pm0.0004$ & $0.231\pm0.0003$ & --- & --- & --- & $1.005$ & 1,625.86 & $88.70\%$ \\    
&&&&&&&&& \\
$\Lambda$CDM & $6.618\pm0.012$ & $0.639\pm0.0005$ & $0.231\pm0.0004$ & $0.31\pm0.05$ & --- & --- & $1.004$ & 1,630.26 & $9.79\%$ \\
&&&&&&&&& \\
Cosmographic& $6.249\pm0.02$ & $0.626\pm0.0006$ & $0.231\pm0.001$ & --- & $2.93\pm0.33$ & $2.65\pm0.80$ & $1.003$ & 1,633.96 & $1.51\%$ \\
&&&&&&&&& \\
\hline\hline
\end{tabular}
\end{center}
\end{table*}

\section{Results}
We have slightly modified the approach of finding $\gamma$ and $\tilde{\beta}$ 
from the method used by Risaliti \& Lusso (2019) but, as we shall see, the 
results are nonetheless highly consistent with theirs. Here, we have avoided 
the use of Type Ia SNe as calibrators, which can be difficult to use for model 
comparisons when various sub-samples are merged together to form a large catalog. 
Even slight differences in the systematics can cause large effects, as discussed 
more thoroughly in Wei et al. (2015) and Melia et al. (2018). A purer test of each 
model using the quasar Hubble diagram bases all of its parameter optimization---those 
of the model itself, in addition to the unknowns $\gamma$, $\tilde{\beta}$,
and $\delta$ appearing in Equations~(1-5)---on the quasar data alone.
This avoids any possible contamination of the calibrators by any
unrecognized model dependence. Using this approach, the reduced data themselves
will change slightly from model to model, since the inferred luminosities---and
therefore the constants $\gamma$, $\tilde{\beta}$, and $\delta$---require the
use of a model-dependent luminosity distance (Eqs.~6-8). This approach is in 
fact analogous to what is done with Type Ia SNe, for which the so-called
`nuisance' parameters characterizing the supernova lightcurve must be
optimized along with the free parameters of the cosmology itself, and therefore
change slightly from model to model. We use this same approach here with the
quasar data, since this appears to be the least biased way of comparing distinct
models, such as $R_{\rm h}=ct$ and $\Lambda$CDM.

\begin{figure}
	\centering
	\includegraphics[scale=0.63]{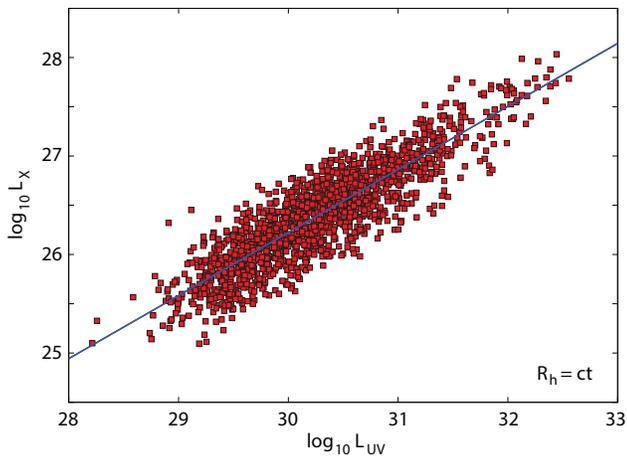}
	\label{Fig.1}\caption{Rest-frame correlation of the monochromatic UV to X-ray
luminosity density for the highly selected 1598 quasar sample. The data and
fit have been optimized for the $R_{\rm h}=ct$ model, which has no free parameters.
A fiducial value $H_0=70$ km s$^{-1}$ Mpc$^{-1}$ has been chosen for the purpose
of display only. The actual value of the Hubble constant does not affect the optimization
of the key parameters, $\gamma=0.64$, $\tilde{\beta}=-13.621$, and $\delta=0.231$.} 
\end{figure}

As we can see in Table~1, we find that the optimized values of the correlation
parameters following this method are $\gamma\approx 0.63$ (for the cosmographic fit) 
and $\approx 0.64$ for both $R_{\rm h}=ct$ and $\Lambda$CDM. For the intrinsic dispersion,
we find a uniform value of $0.231$ in all cases. These are to be compared with
the values $\gamma=0.633\pm0.002$ and $\delta=0.24$ found by Risaliti \& Lusso
(2019) using external calibrators. The values of $\tilde{\beta}$ appearing
in column 1 of this table correspond to our chosen `fiducial' value of the Hubble
constant, i.e., 70 km s$^{-1}$ Mpc$^{-1}$ (see above discussion concerning the
linkage between $\beta$ and $H_0$). And according to Equation~(3), $\beta$ is then
$\sim \tilde{\beta}+0.4$. There are two positive conclusions one may
draw from this comparison: (1) the two approaches, of using or not using external
calibrators, produce results remarkably consistent with each other, and (2) though 
the `nuisance' parameters $\gamma$ and $\delta$ were given the freedom to change 
from model to model, in the end they appear to be only very weakly dependent on 
the presumed cosmology, if at all. This lends support to the idea that the 
correlation assumed in Equation~(1) does in fact provide a reliable standard 
candle for model testing, as we have done in this paper. 

The different methodology we have employed here does produce 
a notable departure from the Risaliti \& Lusso (2019) results, however.
This is most easily recognized in the optimized values of $a_2$ and
$a_3$ for the cosmographic fit, which we find are closer to those
calculated for the standard model by these authors (see fig.~3 in 
Risaliti \& Lusso 2019) whereas, in their case, the cosmographic 
fit differed from $\Lambda$CDM by $\sim \sigma$. Another way to recognized
this is via the fact that all three $\chi^2_{\rm dof}$ shown in Table~1
are very close to each other and close to 1, suggesting that all three
models fit the data very well. This is easily understood in terms of
the internal self-calibration employed in our analysis. Unlike the use
of external calibrators, which essentially fixed the data for Risaliti
\& Lusso, in our case the data must be recalibrated for each model,
so the optimization procedure enhances the quality of the fit for each
cosmology. This is why, in spite of the fact that the 3 curves in fig.~9
are not all the same, particularly at high redshift, each fit is adjudged
to be very good in terms of $\chi^2$, since the data themselves are different
for each model. 

\begin{figure}
        \centering
        \includegraphics[scale=0.63]{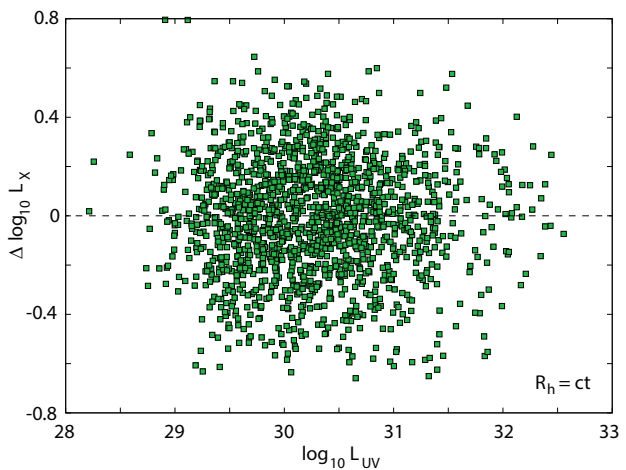}
        \label{Fig.2}\caption{Residuals for the rest-frame monochromatic 
UV to X-ray luminosity density shown in figure~1. The scatter, which is evenly distributed 
in positive and negative values, is reflected in the inferred intrinsic dispersion 
$\delta\sim 0.231$ quoted in Table~1 (see also Eq.~4 and subsequent discussion).}
\end{figure}

\begin{figure}
        \centering
        \includegraphics[scale=0.63]{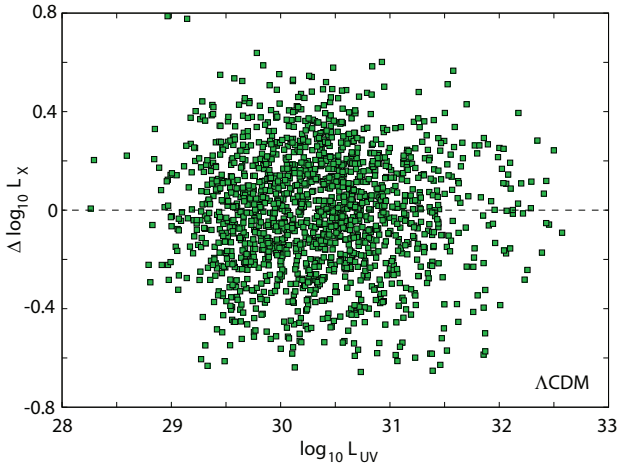}
        \label{Fig.3}\caption{Same as fig.~2, except now for $\Lambda$CDM. The
scatter in these two models is very similar, so their inferred intrinsic dispersions 
$\delta$ are identical (see Table~1).}
\end{figure}

\begin{figure}
        \centering
        \includegraphics[scale=0.63]{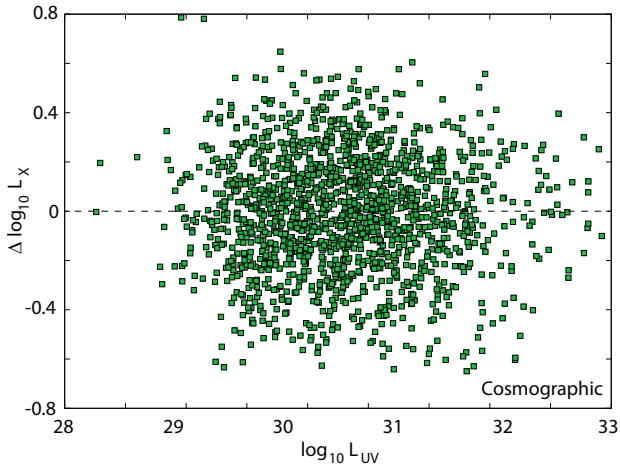}
        \label{Fig.4}\caption{Same as fig.~3, except now for the cosmographic fit.}
\end{figure}

The simultaneous optimization of the `nuisance' and model parameters
resulted in the correlation shown in Figure~1 for the $R_{\rm h}=ct$ universe. The
corresponding figures for $\Lambda$CDM and the cosmographic fit are so similar to
this that there isn't much point in showing all three. Though hardly  perceptable, 
the data do change very slightly from plot to plot, reflecting the fact that
the nuisance parameters are slightly different from one model to the next (Table 1). 
In the end, however, the best-fit correlations are highly consistent among the 
three models we tested here. To complement the fit in figure~1, we also show the 
residuals corresponding to $R_{\rm h}=ct$, $\Lambda$CDM and the cosmographic model 
in figures~2, 3 and 4, respectively. The strong similarity between these model
fits may also be understood by inspecting the curves in figure~9, which shows the
optimized distance moduli. The principal difference between these models is that
the cosmographic fit deviates slightly from the other two at redshifts beyond
$\sim 3$ which, as we shall see, causes a slight bias of the high-redshift Hubble 
diagram residuals to positive values in this model.
 
We then used the optimized parameters to calculate the Hubble diagram in 
each case, which is shown in Figure~5, again for the $R_{\rm h}=ct$
universe. The corresponding figures for $\Lambda$CDM and the cosmographic
model are almost identical, save for slight changes in the recalibrated data. 
From the Hubble diagrams, we determined the quality of the fits (indicated by the 
reduced $\chi^2$ values quoted in Table~1) and calculated the BIC likelihoods. 
Since the sample is so large in this application (specifically $>> 30$), the BIC 
is the preferred information criterion to use for model selection (see, e.g., Melia \&
Maier 2013). As one can see, the reduced $\chi^2$ hardly changes
from one cosmology to the next, reaffirming the view that this measure of 
quality is not sufficient to prioritize the models. The cosmographic
fit actually has a slightly better $\chi^2_{\rm dof}$ than the other
two, but this merely reflects the fact that its greater number of free
parameters gives it greater flexibility to adjust to the noise. 

\begin{figure}
        \centering
        \includegraphics[scale=0.72]{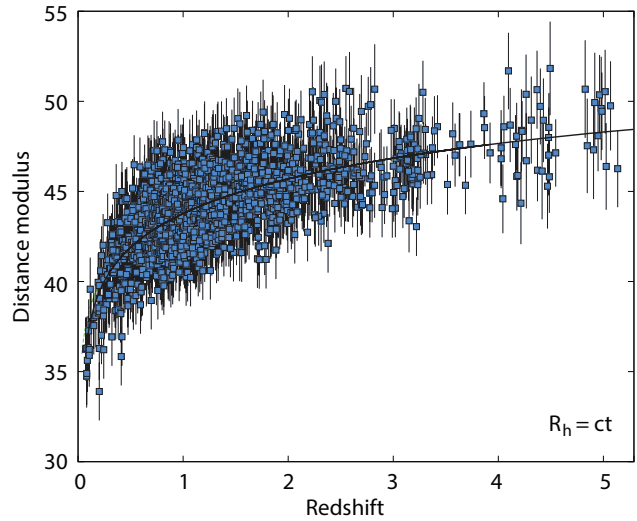}
        \label{Fig.5}\caption{Hubble diagram of the 1598 highly selected quasars for
the $R_{\rm h}=ct$ cosmology. The $\chi^2_{\rm dof}$ for this fit is $1.005$, with
a relative BIC likelihood $\sim 88.7\%$ of this being the correct model.}             
\end{figure}
 
\begin{figure}
        \centering
        \includegraphics[scale=0.72]{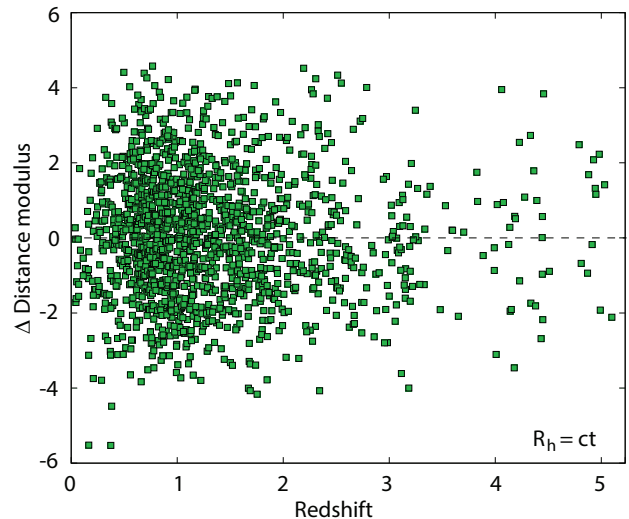}
        \label{Fig.6}\caption{Hubble diagram residuals for the optimized fit shown 
in fig.~5.}
\end{figure}

To complete the comparison, we also show the Hubble diagram
residuals in figures~6, 7 and 8. In spite of the overall similarity between these 
three cosmologies, we do see that, while these residuals are evenly distributed
towards positive and negative values for $R_{\rm h}=ct$ and $\Lambda$CDM, they are 
somewhat biased towards positive values at high redshift for the cosmographic
model. Together with the comparison of all three distance moduli shown in figure~9, 
in which the cosmographic curve is seen to deviate from the other two at high redshift, 
we conclude that the cosmographic ansatz is somewhat disfavoured by these data in 
comparison with both $R_{\rm h}=ct$ and $\Lambda$CDM.

The model prioritization may be put on a firmer quantitative
basis by considering the BIC likelihoods listed in Table~1. Our main conclusion 
in this paper is that $R_{\rm h}=ct$ is preferred by the high-$z$ quasar Hubble
diagram in comparison with both the 1-parameter, flat $\Lambda$CDM cosmology 
and the empirical cosmographic fit. In figure~9 we also learn that $R_{\rm h}=ct$ 
and $\Lambda$CDM are hardly distinguishable from each other, in spite
of the fact that $R_{\rm h}=ct$ has no free parameters for this fit.
The cosmographic fit deviates slightly from the other two at 
$z\gtrsim 3$, but the impact on the quality of the fits is hardly
noticeable. This is due in part to the fact that the data themselves
are recalibrated due to the individual optimization of the nuisance 
variables. 

\begin{figure}
        \centering
        \includegraphics[scale=0.72]{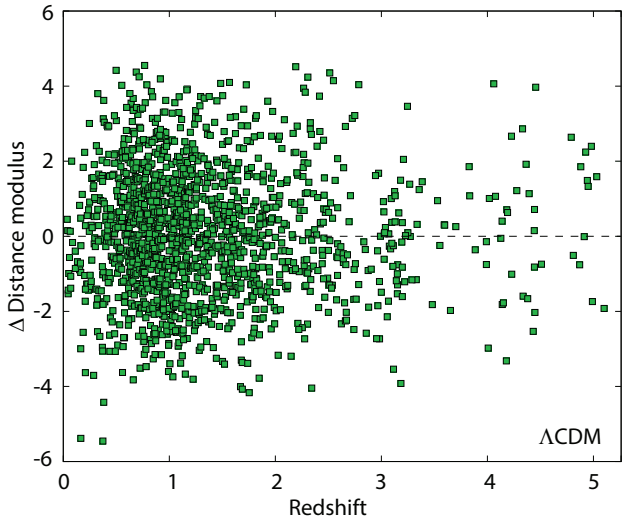}
        \label{Fig.7}\caption{Same as fig.~6, except now for $\Lambda$CDM.}
\end{figure}

\begin{figure}
        \centering
        \includegraphics[scale=0.72]{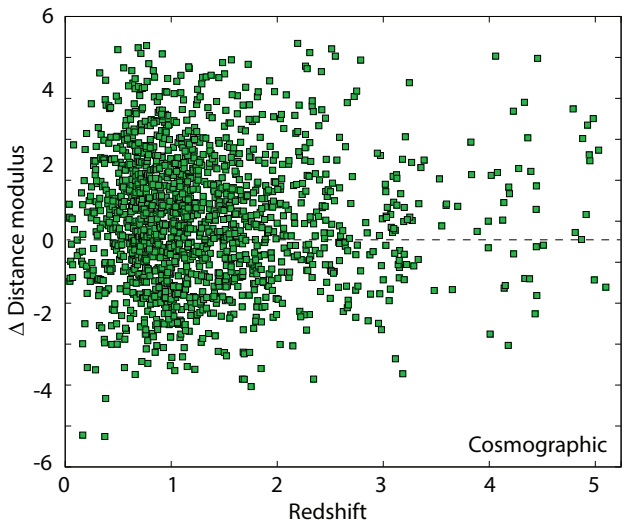}
        \label{Fig.8}\caption{Same as fig.~6, except now for the cosmographic model.}
\end{figure}

It is clear, therefore, that the differences in the BIC
shown in Table~1, and their associated likelihoods, are mostly due
to the different number of adjustable parameters among these three 
models: $R_{\rm h}=ct$ has none, $\Lambda$CDM has one and the
cosmographic fit has two. If we believe that the latter is a 
reliable, model-independent representation of the data, these results 
argue in favour of both $\Lambda$CDM and $R_{\rm h}=ct$ being
validated by the quasar sample. Indeed, both models fit the data
as well as the polynomial expansion in spite of having fewer
parameters. 

This analysis is also permitting us to see---for the first time---how 
the luminosity distances predicted by $R_{\rm h}=ct$ and $\Lambda$CDM 
compare with each other over a very large redshift range. It is 
quite remarkable to see how similar they are over the entire extent
of the observations, $0\lesssim z \lesssim 6$. There is a slight 
variation at $z\sim 2$, but they are virtually indistinguishable 
elsewhere. This result strongly confirms an indication first discussed 
in Melia (2015b), that the predictions of $R_{\rm  h}=ct$ seem to act 
as an `attractor' for the optimization of model parameters in 
$\Lambda$CDM. In other words, the largely empirical parametrization 
in the standard model appears to account very well for many of the 
observations because it has sufficient freedom to mimic the much 
more highly constrained distances in $R_{\rm  h}=ct$. 

\begin{figure}
        \centering
        \includegraphics[scale=0.72]{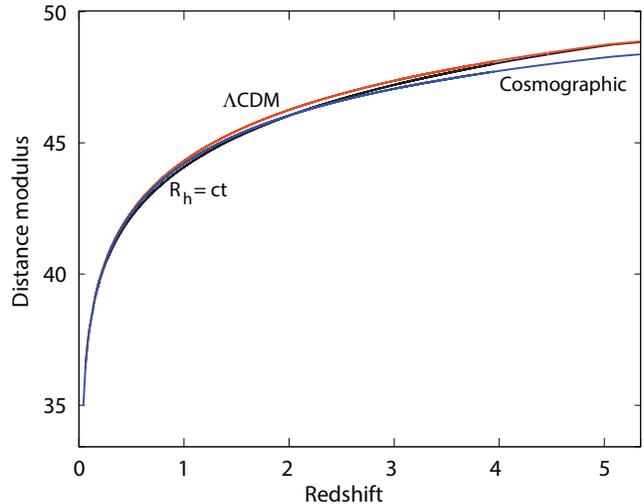}
        \label{Fig.9}\caption{A comparison of all three best-fit 
distance moduli, optimized through fits to the 1598 highly-selected quasars. 
The $\Lambda$CDM curve lies slightly above the other two near $z\sim 2$, while 
the cosmographic curve is the lowest at $z\gtrsim 3$. The $R_{\rm h}=ct$ curve 
corresponds to the fit shown in fig.~5. Though the cosmographic fit starts 
to deviate slightly from the other two at redshifts beyond $\sim 3$, the 
individually re-calibrated data also adjust in concert, with the effect that 
all three reduced $\chi^2$ values are virtually identical. The chief difference 
between these three cases is manifested in the BIC, which also takes into 
account the number of free parameters used in the optimization procedure.}             
\end{figure}

\section{Conclusion}
The outcome of our model comparison using the greatly enhanced high-$z$
quasar sample assembled by Risaliti \& Lusso (2019) has strengthened
earlier results based on the quasar Hubble diagram and FSRQ luminosity
function, showing that $R_{\rm h}=ct$ tends to be favoured by these
data over $\Lambda$CDM. Indeed, an extensive compilation of tests such as these
(see, e.g., Table~2 in Melia 2018a) now shows that $R_{\rm h}=ct$ is a better
match than $\Lambda$CDM to a wide assortment of observations, over a broad
range of redshifts. The analysis in this paper is crucial in extending the
redshift coverage of these comparative studies to intermediate redshifts
($2\lesssim z\lesssim 6$), where few other sources are readily available.

At a redshift $z\sim 6$, the Universe was about $1.9$ Gyrs old in $R_{\rm h}=ct$
(and roughly $900$ Myr in $\Lambda$CDM), so the work reported here covers a
sizeable fraction of the visible Universe. We have found that for much of the
evolutionary history in the cosmos, the luminosity distance predicted by the 
$R_{\rm h}=ct$ model, which has no free parameters for the work reported
here, acts as an `attractor' for the optimization of model parameters in 
$\Lambda$CDM. This follows a trend seen earlier using other kinds of data,
including Type Ia SNe and anisotropies in the cosmic microwave background
(Melia 2015b), all of which resulted in optimized fits with $\Lambda$CDM
that mimicked the predictions in $R_{\rm h}=ct$, whose tightly constrained
observational signatures lack the flexibility of freely adjusting to `noise'
in the data.  

An important by-product of our analysis is the affirmation that the correlation 
seen between the X-ray and UV monochromatic luminosities in these high-$z$
quasars does reliably produce a standard candle for cosmological testing at 
redshifts not easily accessible by other means, e.g., Type Ia SNe. We have
found that the optimization of the `nuisance' parameters characterizing this
correlation depends only weakly on the chosen cosmology. In doing so, we
have demonstrated that a reasonable approach in tests such as these is to
optimize all of the parameters together---those of the model itself and those 
in the correlation function---without resorting to external calibrators to
fix any of them. One may therefore rely on a high quality sample of quasars
such as this for the entire comparative analysis without the concern of
introducing possible biases from unrelated data. 

As the $R_{\rm h}=ct$ cosmology continues to be validated with tests such
this, especially towards higher redshifts, its viability strengthens the
view that the Universe may actually not have a horizon problem, either in
temperature (Melia 2013b), or the electroweak phase transition (Melia 2018b),
thereby relinquishing the need of developing a self-consistent inflationary
paradigm. Even after four decades of attempts at doing so, we still lack
a complete theory for how and when inflation could have taken place. In
some sense, one may argue that evidence is now accumulating disfavouring
the basic inflationary concept (see, e.g., Ijjas et al. 2013, 2014). 

A recent measurement of the minimum cutoff $k_{\rm min}$ seen in the power
spectrum of the cosmic microwave background appears to be incompatible
with the basic premise of quantum fluctuations crisscrossing the horizon
during an inflationary phase, suggesting that one may not simultaneously
solve the horizon problem and the formation of structure in the standard
model with the same phase transition (Melia \& L\'opez-Corredoira 2018; 
Liu \& Melia 2019). The work reported here lends some support to the 
view that inflation may not be needed, and perhaps never happened.

\section*{Acknowledgments} I am grateful to the anonymous
referees for comments that have lead to a significant improvement in the 
presentation of these results. I am especially grateful to Beta Lusso 
and Guido Risaliti for sharing their data for this analysis. I am also
grateful to the Instituto de Astrof\'isica de Canarias in Tenerife and 
to Purple Mountain Observatory in Nanjing, China for their hospitality 
while part of this research was carried out.

\label{lastpage}

\begin{thebibliography}{99}
\bibitem{Akaike:1973} Akaike, H., Petrov, B. N. \& Csaki, F., 1973,
         Second International Symposium on Information Theory, Budapest Akademiai Kiado p. 267
\bibitem{Avni:1986} Avni, Y. \& Tananbaum, H., 1986, ApJ, 305, 83
\bibitem{Banados:2014} Banados, E. et al., 2014, AJ, 148, 14
\bibitem{Bentz:2009} Bentz, M. C., Peterson, B. M., Netzer, H., Pogge, R. W. \& Vestergaard, M.,
        2009, ApJ 697, 160 
\bibitem{Blandford:1982} Blandford, R. D. \& McKee, C. F., 1982, ApJ, 255, 419
\bibitem{Fan:2003} Fan, X. et al., 2003, AJ, 125, 1649
\bibitem{Jiang:2007} Jiang, L., Fan, X., Vestergaard, M., Jurk, J. D., Walter, F.,
         Kelly, B. C., Strauss, M. A., 2007, AJ, 134, 1150
\bibitem{Jones:2013} Jones, D. O. et al., 2013, ApJ, 768, id. 166
\bibitem{Just:2007} Just, D. W. et al., 2007, ApJ, 665, 1004
\bibitem{Ijjas:2013} Ijjas, A., Steinhardt, P. J. \& Loeb, A., 2013, PLB, 723, 261
\bibitem{Ijjas:2014} Ijjas, A., Steinhardt, P. J. \& Loeb, A., 2014, PLB, 736, 142
\bibitem{Kaspi:2000} Kaspi, S., Smith, P. S., Netzer, H., Maoz, D., Jannuzi, B. T. \& Giveon, U.,
        2000, ApJ, 533, 631
\bibitem{Kass:1995} Kass, R. E. \& Raftery, A. E. 1995, J. Amer. Statist. Assoc., 90, 773
\bibitem{Kim:2011} Kim, A. G., 2011, Publ. Astron. Soc. Pac., 123, 230
\bibitem{Lin:2018} Lin, H.-N., Li, X. \& Sang, Y., 2018, Chinese Physics C, 42, 095101
\bibitem{Liu:2019} Liu, J. \& Melia, F., 2019, PRL, submitted
\bibitem{Lopez;2018} L\'opez-Corredoira, M., Melia, F., Lusso, E. \& Risaliti, G., 2018, 
IJMP-D, 25, id. 1650060
\bibitem{Lusso:2010} Lusso, E. A. et al., 2010, A\&A, 512, A34
\bibitem{Lusso:2016} Lusso, E. A. \& Risaliti, G., 2016, ApJ, 819, 154
\bibitem{Melia:2003} Melia, F., 2003, ``The Edge of Infinity: Supermassive Black Holes in the Universe,"
         Cambridge University Press, Cambridge
\bibitem{Melia:2007} Melia, F., 2007, MNRAS, 382, 1917
\bibitem{Melia:2012} Melia, F. \& Shevchuk, A.S.H., 2012, MNRAS, 419, 2579
\bibitem{Melia:2013a} Melia, F., 2013a, ApJ, 764, 72
\bibitem{Melia:2013b} Melia, F., 2013b, A\&A, 553, A76
\bibitem{Melia:2014} Melia, F., 2014, JCAP, 01, 027
\bibitem{Melia:2015a} Melia, F., 2015a, Astrophys. Sp. Sci., 359, 34
\bibitem{Melia:2015b} Melia, F., 2015b, Astrophys. Sp. Sci., 356, 393
\bibitem{Melia:2017} Melia, F. \& L\'opez-Corredoira, M., 2017, IJMP-D, 26, id. 1750055-265
\bibitem{Melia:2018a} Melia, F., 2018a, MNRAS, 481, 4855
\bibitem{Melia:2018b} Melia, F., 2018b, EPJ-C Lett., 78, 739
\bibitem{Melia:2018d} Melia, F. \& L\'opez-Corredoira, M., 2018, A\&A, 610, A87
\bibitem{Melia:2018e} Melia, F. et al. 2018e, EPL, 123, 59002
\bibitem{MeliaMaier:2013} Melia, F. \& Maier, R. S., 2013, MNRAS, 432, 2669
\bibitem{Melia:2015c} Melia, F. \& McClintock, T. M., 2015, Proc. R. Soc. A, 471, 20150449
\bibitem{Melia:2018c} Melia, F., Wei, J.-J., Maier, R. S. \& Wu, X., 2018, EPL, 123, 59002
\bibitem{Mortlock:2011} Mortlock, D. J. et al., 2011, Nature, 474, 616
\bibitem{Paris:2017} Paris, I. et al., 2017, A\&A, 597, A79
\bibitem{Risaliti:2015} Risaliti, G. \& Lusso, E. A., 2015, ApJ, 815, 33
\bibitem{Risaliti:2019} Risaliti, G. \& Lusso, E. A., 2019, Nature Astronomy, in press
\bibitem{Rosen:2016} Rosen, S. R. et al., 2016, A\&A, 590, A1
\bibitem{Schwarz:1978} Schwarz, G., 1978, Ann. Stat., 6, 461
\bibitem{Shafer:2015} Shafer, D. L., 2015, PRD, 91, 103516
\bibitem{Shen:2008} Shen, Y., Greene, J. E. \& Strauss, M. A., 2008, ApJ, 680, 169
\bibitem{Shen:2011} Shen, Y. et al., 2011, ApJ Supp, 194, 45
\bibitem{Steinhardt:2010} Steinhardt, C. L. \& Elvis, M., 2010, MNRAS Lett., 406, L1
\bibitem{Tan:2012} Tan, M. Y. J. \& Biswas, R. 2012, MNRAS, 419, 3292
\bibitem{Volonteri:2006} Volonteri, M. \& Rees, M. J., 2006, ApJ, 650, 669
\bibitem{Wandel:1999} Wandel, A., Peterson, B. \& Malkan, M., 1999, ApJ, 526, 579
\bibitem{Wei:2015} Wei, J.-J., Wu, X., Melia, F. \& Maier, R. S., 2015, AJ, 149, 102
\bibitem{Willott:2007} Willott, C. J. et al., 2007, AJ, 134, 2435
\bibitem{Willott:2010} Willott, C. J. et al., 2010, AJ, 140, 546
\bibitem{Yoo:2004} Yoo, J. \& Miralda-Escud\'e, J. 2004, ApJ, 614, L24
\bibitem{Young:2010} Young, M., Risaliti, G. \& Elvis, M., 2010, ApJ, 708, 1388
\bibitem{Zeng:2016} Zeng, H., Melia, F. \& Zhang, L., 2016, MNRAS, 462, 3094

\end{thebibliography}
\end{document}